# Emulsion-templated formation of poly(N-isopropylacrylamide):surfactant mixed shells by thermo-enhanced interfacial complexation

Lucas Sixdenier*, Christophe Tribet, Emmanuelle Marie*

*PASTEUR, Département de chimie, École normale supérieure, PSL University, Sorbonne Université, CNRS, 75005 Paris, France*

*Corresponding authors.

E-mail addresses: lucas.sixdenier@ens.psl.eu, emmanuelle.marie@ens.psl.eu.

# Abstract

The encapsulation of fragile biomacromolecules is crucial for many biotechnological applications but remains challenging. Interfacial complexation (IC) in water-in-oil emulsions turned out to be an efficient process for the formation of protective polymer layers at the surface of capsule-precursor water droplets. We propose to further enhance this IC process by introducing thermoresponsive poly(N-isopropylacrylamide) (PNIPAM) strands in the interfacial polymer layer. In this work, we implement surfactant-polymer IC in water-in-fluorocarbon oil emulsions between a water-soluble poly(L-lysine)-g-poly(N-isopropylacrylamide) cationic copolymer (PLL-g-PNIPAM) and an oil-soluble anionic surfactant. We demonstrate that the thermal collapse transition of PNIPAM strands triggers an enrichment of the polymer layer initially formed by IC. This process is leveraged to irreversibly segregate water-soluble nanoparticles in the interfacial polymer layer, resulting in gel-like mixed shells. We demonstrate that this thermo-enhancement of conventional IC is a promising approach for the formation, strengthening and functionalization of capsule shells. As implemented in mild conditions, thermo-enhanced IC is additionally compatible with the encapsulation of proteins, paving the way for new delivery systems of biomacromolecules.

# 1. Introduction

Compartmentalization of molecules or nanoparticles is commonly required for various biotechnological applications including drug encapsulation and delivery[1], cell mimicry[2] or molecular screening and analysis[3]. Encapsulation of biomacromolecules such as RNA or proteins is particularly demanding because of their fragility and high sensitivity to their local environment[4]. Achieving efficient encapsulation without altering the integrity and activity of proteins requires specifically to preserve a mild aqueous environment in artificial compartments[5]. Capsular structures have been developed to confine biomolecules within protective shells, avoiding contacts with external apolar solvents or low or high pHs that may be denaturing. In such capsular systems, water-insoluble polymers are typically used to form the external shell wall, allowing not only to control the shell structure (thickness, porosity, rigidity, etc.)[6] but also to impart functionalities to the capsule (stimuli-responsiveness, chemical activity, specific recognition, etc.)[7]. In addition to polymer chains the capsule shell may also contain organic or inorganic particles that may increase mechanical properties and/or introduce new functionalities[8].

The formation of polymer capsules usually requires a template to guide the assembly of macromolecular chains into shells with well-defined sizes and morphologies. The various procedures used for template-guided shell formation − interfacial polymerization[9], internal phase separation[10], layer-by-layer assembly[11] − may however be stressful and of limited interest for the encapsulation of biomolecules. For instance, interfacial polymerization and internal phase separation are usually implemented with oil-based cores and with solvents of polymers that may denature proteins[12]. Layer-by-layer assembly most of the time requires sacrificial solid templates that must be removed in harsh conditions inappropriate to fragile biomolecules[13].

In contrast, interfacial complexation proved to be an easy and versatile approach for the formulation of polymer envelopes delimiting aqueous cores in mild conditions. In this process, two oppositely charged polymers are solubilized in each phase of an emulsion, and spontaneously form a complex at the liquid-liquid interface via attractive electrostatic interactions. Interfacial complexation of polyelectrolytes has been achieved in water-in-oil[14,15], water-in-water[16,17] or water-in-water-in-water[18,19] emulsions. All-aqueous emulsions have the advantage to readily produce water-compatible capsules useable as cargos for biological applications. Proteins or cells have been successfully entrapped in all-aqueous assemblies but the partitioning of water-soluble species between inner and outer phases may limit the encapsulation efficiency[20]. Improving the yield of encapsulation of fragile proteins in mild conditions remains challenging, especially for biological applications that require high molecular concentrations and high activity preservation.

The use of fluorocarbon oils (FCO) as continuous phase of water-in-oil emulsions turned out to be an interesting alternative to hydrocarbon oils since solution components (salts, hydrocarbon species, biomolecules) show an extremely poor solubility in FCO, offering an improved control on the composition of the aqueous droplets. To our knowledge, very few works have implemented surfactant-polymer interfacial complexation (SPIC) in water-in-FCO emulsions for the design of capsular systems. DeJournette et al. have evidenced the formation of a complex layer by SPIC between an FCO-soluble macrosurfactant (a carboxylate-terminated perfluoro-PEG called Krytox) and a water-soluble diamine (Jeffamine) at the edge of water droplets[21]. More recently, Abell and co-workers showed that SPIC can be leveraged to locally increase the concentration of a polycation at the water/FCO interface and implemented supramolecular cross-linking of adjacent chains by cucurbituril units to form thin polymer capsules[22,23].

Based on SPIC in water-in-FCO (W/FCO) emulsion, we propose here a general method to modulate the deposition of polymer chains at the water/FCO interface, enabling (i) the formation and enrichment of a polymer shell, (ii) the straightforward incorporation of nanoparticles and (iii) the encapsulation of proteins in mild conditions. Our strategy relies on the presence of temperature-responsive strands in the water-soluble polymer, enabling the amplification of the SPIC process by additional temperature-mediated accumulation of polymer chains at droplet interfaces. We show that a germinal polymer layer was formed at ambient temperature − i.e. in conditions of conventional SPIC − between Krytox surfactant in the FCO continuous phase and poly(L-lysine)-g-poly(N-isopropylacrylamide) copolymer (PLL-g-PNIPAM) at the edge of aqueous droplets. PNIPAM is a thermoresponsive polymer exhibiting a hydrophilic-to-hydrophobic transition, switching from swollen coil to collapsed globule, at a Lower Critical Solution Temperature of 32-34°C in water. Accordingly, grafting PNIPAM strands onto PLL introduced temperature-sensitivity in the polymer layer formed at the W/FCO interface. We show that the thermal collapse transition of PNIPAM significantly increases the amount of polymer chains deposited in the interfacial layer initially self-organized by SPIC. We demonstrate that this thermo-enhanced process can be leveraged to form polymer-particles mixed shells by capture of nanoparticles in the interfacial polymer layer. This novel and simple approach for the formation of polymer envelopes, combining interfacial complexation and thermo-responsiveness of PLL-g-PNIPAM, was achieved upon gentle heating and in mild pH conditions, turning out to be compatible with the encapsulation of a model protein and paving the way for promising functional encapsulation and delivery systems.

## 2. Materials and Methods

*2.1. Materials*

Poly(L-lysine) hydrobromide (PLL, MW = 15-30 kDa), (N-hydroxysuccinimidyl ester)-terminated poly(N-isopropylacrylamide) (PNIPAM-NHS, MW = 2 kDa), azide-terminated poly(N-isopropylacrylamide) (PNIPAM-$N_3$, MW = 5 kDa), 5(6)-carboxy-X-rhodamine N-hydroxysuccinimidyl ester (rho-NHS), 7-(diethylamino)coumarin-3-carboxylic acid N-succinimidyl ester (coum-NHS), dibenzocyclooctyne-Cy3 (DBCO-Cy3) and 8-anilino-1-naphthalenesulfonic acid (ANS) were purchased from Sigma-Aldrich. (N-hydroxysuccinimidyl ester)-terminated poly(ethylene glycol) (PEG-NHS, MW = 2 kDa) was purchased from Rapp Polymere. Fluorinert FC-70 fluorocarbon oil (FCO) was purchased from Sigma-Aldrich. Krytox 157 FS(L) (MW = 2.5 kDa) was purchased from Samaro. FluoSpheres™ NeutrAvidin™-Labeled Microspheres (40 nm diameter, 1 %wt in pure water) displaying yellow-green fluorescence (505/515 nm) were purchased from ThermoFisher Scientific. Unless otherwise noticed, all chemicals were used without further purification. For the preparation of aqueous solutions, water-soluble species were systematically dissolved or diluted in a home-made phosphate buffer, prepared by dissolving 360 mg of $NaH_2PO_4$ (3.0 mmol) and 670 mg of $Na_2HPO_4$ (4.7 mmol) in 1L of DI water (pH = 7.3, ionic strength = 17 mM, Debye length = 2.3 nm). Enhanced Green Fluorescent Protein (eGFP) has been produced and purified according to a protocol described in [24].

*2.2. Polymer synthesis and functionalization*

PLL was grafted with NHS-terminated strands (PNIPAM-NHS or PEG-NHS) and labeled with NHS-terminated fluorophores (rho-NHS or coum-NHS). The amount of NHS-terminated strands was adjusted as a molar fraction of total hydrobromide lysine units (M = 209 g/mol) to reach the targeted grafting ratio τ (number of functionalized lysine units/total number of lysine units).

Here is a typical protocol for the synthesis of rhodamine-labeled PLL-g-PNIPAM: PLL hydrobromide (MW = 15-30 kDa, 15 mg) was dissolved in 5 mL of sodium tetraborate buffer (pH = 8.5). PNIPAM-NHS (MW = 2 kDa, 48 mg for τ = 0.33) and rho-NHS (430 µL of rho-NHS at 1 g/L solution in DMSO for τ = 0.01) were then added and the solution was stirred at 4°C until complete dissolution of the polymer. Then, the solution was stirred for an additional 4h at room temperature. The resulting solution was dialyzed against DI water for 3 days in a Thermo Scientific Slide-A-Lyser cassette (3-12 mL, MW cutoff = 3.5 kDa) and finally freeze-dried for 2 days in a Labconco Freezone Plus 2.5 apparatus, yielding a colored fluffy powder (144 mg, yield = 83%). $^1$H NMR (400 MHz, D$_2$O, δ in ppm): 1.0-1.3 (lysine γ-CH2 + PNIPAM [CH$_2$-CH-CO-NH-CH-(C**H**$_3$)$_2$], 46H), 1.3-2.2 (lysine β,δ-CH$_2$ + PNIPAM [C**H**$_2$-C**H**-CO-NH-CH-(CH$_3$)$_2$], 31H), 2.8-2.9 (non-grafted lysine α-CH$_2$, 1.76H), 3.20 (grafted lysine α-CH$_2$, 0.41H), 3.90 (PNIPAM [CH$_2$-CH-CO-NH-C**H**-(CH$_3$)$_2$], 7.6H), 4.30 (lysine backbone, 1H). The grafting ratio τ of PLL by polymer strands was determined by comparing $^1$H-NMR integration values of protons characterizing non-grafted (1.76H) and grafted (0.41H) lysine units, with protons of PLL backbone as reference protons (1H).

The functionalization of PNIPAM-N$_3$ by DBCO-Cy3 was performed as follows: PNIPAM-N$_3$ (MW = 5 kDa, 100 mg) was dissolved in 5 mL of DI water. DBCO-Cy3 (23 µL at 1 g/L in PBS buffer) was added and the solution was stirred for 5 h at room temperature.

The resulting solution was dialyzed against DI water for 3 days (Slide-A-Lyser cassette, MW cutoff = 3.5 kDa) and finally freeze-dried for 2 days, yielding a pink fluffy powder.

*2.3. Turbidimetry by UV-Visible spectrophotometry*

UV-Visible spectra were recorded on a single cell Thermo Scientific Evolution Array UV-Vis spectrophotometer equipped with a Peltier temperature-controlled cell holder (+/- 0.1°C). Sample solutions were injected in a 60 µL quartz micro-cuvette (OPL = 3 mm) and submitted to an increase of temperature from 20°C to 60°C by steps of 2°C. A spectrum was recorded at every step temperature after 3 min thermalization and the mean transmittance between 650 and 750 nm was used to plot the turbidimetry curve.

*2.4. Emulsion formulation and observation*

5 vol% water-in-FCO emulsions were prepared as follows: 5 µL of aqueous phase (containing the polymer at 1-5 wt% in phosphate buffer, pH = 7.3) and 95 µL of FCO phase (containing 0.01-0.05 wt% Krytox) were mixed in a 0.5 mL Eppendorf tube and manually emulsified until the mixture became turbid. For microscopy experiments, a few µL of emulsion were injected in a hollow rectangle capillary (H x W = 0.1 x 1 mm, CMScientific) previously coated with PLL-g-PEG to prevent the breakage of water droplets at the contact with bare glass side. The internal sides of capillaries were coated as follows: capillaries were filled with an aqueous solution of PLL-g-PEG (1 g/L in phosphate buffer) and incubated for 30 min before evaporation on a heating plate at 60°C. Then DI water was introduced by capillarity and evaporated at 60°C.

Phase contrast imaging was performed with a LEICA DM IRE2 microscope equipped with a long-focal ×63 air objective. Images were acquired with a Retina 6000 Q-imaging camera and processed with Micro-Manager 1.4 software (Image J).

*2.5. Confocal Laser Scanning Microscopy*

Confocal images were acquired on a Zeiss LSM 710 META Laser Scanning Microscope equipped with a Plan Apochromat x20 (0.8 NA) air objective and a temperature-controlled heating stage (25-50°C). To ensure emulsion thermalization, capillaries filled with the emulsion were placed on a pierced supporting copper slide that was fixed on the heating stage of the confocal microscope. Low temperatures (16-20°C) were achieved by placing a glass petri dish full of ice on top of the supporting copper slide close to the capillary. The actual temperature of the supporting copper slide was measured with a REED ST-640B thermocouple. The excitation of ANS and coumarin (both at 405 nm) was performed using a 30mW CW/pulsed diode laser. The excitation of eGFP and FluoSpheres™ NeutrAvidin™-Labeled Microspheres (both at 458 nm), rhodamine and Cy3 (both at 514 nm) was performed using a 25 mW argon laser. Images were acquired using LSM ZEN 2009 software and processed with Fiji software (ImageJ).

# 3. Results and discussion

*3.1. Formulation of emulsions and interfacial complexation of PLL derivatives*

Poly(L-lysine)-g-poly(N-isopropylacrylamide) copolymer (PLL-g-PNIPAM) was synthesized by grafting reactive PNIPAM strands (NHS-terminated, MW = 2000 g/mol) onto a PLL backbone (MW = 15-30 kDa) with a grafting ratio of ~ 0.20 (see Materials and Methods and [25]). PLL-g-PNIPAM was labeled with a fluorophore (coumarin or rhodamine) to evaluate by fluorescence microscopy the efficiency of surfactant-polymer interfacial complexation (SPIC) between PLL-g-PNIPAM (polymer) and Krytox (surfactant) in water-in-fluorocarbon oil (W/FCO) emulsion. Emulsions containing free PNIPAM chains (labeled with Cy3) were used as a control since PNIPAM was not expected to accumulate at the interface if not associated to cationic PLL (**Fig. 1.A**). W/FCO emulsions were prepared as follows: the continuous oil phase was made of FC-70 fluorocarbon oil with varying concentrations of Krytox. An aqueous solution of PLL-g-PNIPAM (in the %wt range) was manually emulsified in the FCO solution in a 5:95 volume ratio, yielding nano- to micrometer size water droplets dispersed in the continuous oil phase. This W/FCO emulsion was injected in a glass capillary for microscopy imaging by confocal laser scanning microscopy (CLSM).

At 3 wt% of rhodamine-labeled PLL-g-PNIPAM and 0.05 wt% Krytox, the fluorescence of rhodamine was essentially localized at the edge of the water droplets, i.e. at the water/FCO interface, as shown in **Fig. 1.B**. PNIPAM free chains (in absence of PLL) were in contrast homogenously distributed in the core of the water droplets (**Fig. 1.B**). The difference of distribution between free PNIPAM and PLL-g-PNIPAM is highlighted in the radial fluorescence profiles (**Fig. 1.C**), with a sharp peak localized at the droplet edge for PLL-g-PNIPAM and a uniform level of fluorescence within the droplet core for free PNIPAM. The higher density of PLL-g-PNIPAM at the interfaces (as compared to both the droplet cores and the continuous oil phase) combined with a high emulsion stability (absence of droplet coalescence for 5 hours, see **Fig. S1**) are indicative of the formation of an interfacial polymer layer by SPIC when PNIPAM is grafted onto the cationic PLL backbone.

For each batch of PLL derivatives, the Krytox/PLL-g-PNIPAM concentration ratio was adjusted to achieve an optimal deposition of PLL-g-PNIPAM at the interface by SPIC, evidenced by (i) the stability of the emulsion and (ii) the formation of a fluorescent layer of PLL-g-PNIPAM at the droplet edge. This optimization is illustrated in **Fig. 2** in the case of a coumarin-labeled PLL-g-PNIPAM. At low Krytox concentration (typically $\leq 0.01$ wt%) the coalescence of droplets onto the capillary side revealed that the surfactant concentration was insufficient to stabilize the water droplets. At high Krytox concentration (typically $\geq 0.05$ wt%), the fluorescence of coumarin-labeled PLL-g-PNIPAM was essentially observed in the continuous phase, indicating a massive extraction of polymer chains in the FCO. As the FCO does not solubilize the hydrophilic hydrocarbon-based PLL derivatives, this suggests that excess Krytox can disperse hydrophilic compounds in FCO, probably under the form of inverse micelles[26,27]. At intermediate wt% of Krytox (typically 0.03 wt%), fluorescence micrographs showed a bright corona of PLL-g-PNIPAM localized at the droplet edges and negligible fluorescence in the oil. This is consistent with a SPIC process where the PLL-g-PNIPAM copolymer is attracted towards the interface upon the formation of coulombic complexes with Krytox as schematized in **Fig. 1**. At high amount of polymer in the aqueous phase (typically $\geq 5$ wt%), the interface was saturated in PLL-g-PNIPAM and excess polymer chains remained soluble within the droplet core, leading to a low fluorescence contrast between the edge and the core of the droplets. Because of the excessive amount of internal polymer content, this condition was not suitable for the formation of hollow polymer capsules.

*3.2. Thermoresponsive behavior of the interfacial PLL-g-PNIPAM layer*

The thermoresponsive behavior of PLL-g-PNIPAM was characterized in solution by turbidimetry and compared to free PNIPAM (**Fig 3.A**). The transmittance profiles of either PLL-g-PNIPAM or free PNIPAM solutions decreased between 33°C and 40°C, indicating that PNIPAM strands – grafted or not – undergo a coil-to-globule transition in the same temperature window. A linear extrapolation of the curve around the inflexion point indicates the same transition temperature of 34°C for both PNIPAM forms (coinciding with the beginning of the decrease in transmittance). In the following, this temperature of collapse transition of PNIPAM strands will be referenced as $T_c$.

To assess the thermal transition of PLL-g-PNIPAM in conditions of SPIC with Krytox (in W/FCO droplets), we used 8-anilinonaphthalene-1-sulfonic acid (ANS) as a fluorescence probe exhibiting enhanced fluorescence signal when confined in hydrophobic regions[28]. ANS fluorescence was used as a marker of (i) PLL-g-PNIPAM localization since polymer-enriched regions are more hydrophobic than the aqueous background of water droplets and (ii) thermoresponsive properties of PNIPAM since its collapse transition to a more hydrophobic state would be evidenced by an increase in ANS fluorescence intensity. **Fig. 3.B** shows confocal micrographs of a W/FCO emulsion with 2 wt% of PLL-g-PNIPAM and 500 µM of ANS in water at T = 20°C and T = 45°C, respectively below and above $T_c$ = 34°C.

At T = 20°C the fluorescence signal of ANS in polymer-loaded droplets was non uniform: the intensity was higher at the droplet edge, as highlighted by the sharp peak of fluorescence (~ 3-fold higher than the core level) observed in the radial profile (**Fig. 3.C**). This observation, suggesting that the droplet edges are more hydrophobic than the droplet cores, is consistent with an initial deposition of PLL-g-PNIPAM at the water/FCO interface as shown in **Fig. 1**. As temperature was increased from 20°C to 45°C, we observed an increase in fluorescence contrast between the edge and the core of the droplets (**Fig. 3.B**), due to a concomitant decrease of intensity in the core and increase of intensity of the peak at the edge (**Fig. 3.C**). The ratio between edge and core fluorescence intensities plotted in **Fig. 3.D** shows a continuous increase with temperature, with a significant change of slope at T = 34°C corresponding to $T_c$. The significant values of the edge/core ratio at high temperature (T > $T_c$) confirm the occurrence of the collapse thermal transition of PNIPAM at the water/FCO interface at T = $T_c$. Moreover, the fluorescence level of ANS fell down to almost zero in droplet cores at T > $T_c$, suggesting that PLL-g-PNIPAM chains that were initially soluble in the droplet cores have enriched the interface upon heating.

The reversibility of this process was assessed by submitting an emulsion to temperature cycles while recording the edge/core ratio of ANS fluorescence (**Fig. S2**): initially set at T = 18°C, the emulsion was heated up to 45°C, then cooled back to 18°C and heated again to 45°C. In samples brought back to T = 18°C, the fluorescence level in the droplet cores slightly increased compared to T = 45°C but did not come back to the initial value measured before heating. After the second heating step the fluorescence profile was identical to the one after the first heating step. These observations point out that after a first collapse transition in the SPIC layer, PLL-g-PNIPAM that had accumulated at the droplet edge remained stuck in the interfacial layer and did not significantly redissolve when the sample was brought back below the transition temperature (at least on experimental time scale of ~ 1 hour).

As a control experiment, the same imaging process was performed with a non-thermoresponsive poly(L-lysine)-g-poly(ethylene glycol) (PLL-g-PEG) copolymer with the same 0.20 grafting ratio than PLL-g-PNIPAM. A solution of 2 wt% of PLL-g-PEG and 500μM of ANS was emulsified in FCO and the emulsion was observed by CLSM at T = 18°C and T = 45°C (**Fig. S3**). After formulation at T = 18°C, the droplet edge/core ratio of ANS fluorescence was greater than 1, indicating that PLL-g-PEG has been adsorbed at the water/oil interface by SPIC with Krytox, but significantly lower than for PLL-g-PNIPAM, probably because both PLL and PEG moieties are highly hydrophilic. Moreover, no significant change of the edge/core ratio in droplets containing PLL-g-PEG was observed between T = 18°C and T = 45°C, indicating that temperature had no effect on the hydrophobicity of the interfacial polymer layer. These results confirm that the thermo-responsiveness of PLL-g-PNIPAM highlighted by ANS fluorescence at the droplet edges is intrinsically related to the collapse transition of PNIPAM strands.

*3.3. Effect of droplet radius and temperature on characteristic features of the interfacial layer.*

To better characterize the interfacial polymer layer, we implemented a quantitative approach relying on fluorescence imaging. As the fluorescence of carboxy-X-rhodamine (rho) is neither subject to bleaching [29] nor sensitive to temperature (**Fig. S4**), it was used to evaluate the concentration of rho-labeled PLL-g-PNIPAM (PLLrho-g-PNIPAM) in emulsion over temperature changes. **Fig. 4.A** shows confocal micrographs of a W/FCO emulsion with water droplets containing 3 wt% of PLLrho-g-PNIPAM at T = 18°C and T = 45°C. As previously showed in **Fig. 1**, the fluorescence of rho was brighter at the droplet edge, indicating that PLLrho-g-PNIPAM had accumulated at the water/FCO interface by conventional SPIC upon the emulsification process. At T = 45°C (above $T_c$), the interface was still highly fluorescent, but the fluorescence signal had almost disappeared in the droplet core. This sharp decrease in core intensity upon heating is highlighted in the radial profiles drawn in **Fig. 4.A**. This observation supports the idea of a thermo-induced transfer of core-soluble PLLrho-g-PNIPAM chains towards the interface, probably related to the progressive immobilization of diffusive chains onto the excess interfacial PLLrho-g-PNIPAM layer via hydrophobic interactions. This hypothesized process is depicted in **Fig. 4.B**.

The "apparent" thickness of the interfacial polymer layer – corresponding to the width at half height of the fluorescence peak at the droplet edge – had similar values of ~ 1.5 µm for different polymer concentrations (in the 2-5 wt% range) and at low and high temperatures (**Fig. S5**). However, lack of resolution of the fluorescence together with optical bias affecting the trajectory of light at droplet interfaces did not enable to precisely extract the layer thickness.

To achieve robust and reliable quantification of the amount of PLLrho-g-PNIPAM in the interfacial layer, we measured the concentration of the polymer soluble fraction (in the core) and deduced, by difference to the initial composition, the quantity of polymer adsorbed at the interface. Raw values of rho fluorescence were however biased by optical effect of droplet curvature since the water/FCO interface can deviate the light beam of the microscope. To account for optical bias on fluorescence within a droplet, the core fluorescence of PLLrho-g-PNIPAM was normalized following a suitable calibration process (described in Supporting Information, **Fig. S6**). Briefly, this calibration included a correction factor determined by measuring the fluorescence intensity of a soluble Cy3-labeled PNIPAM that was uniformly distributed within droplets (as previously shown in **Fig. 1**) as a function of droplet diameter from 10 to 50 µm (**Fig. S6**). Once the actual PLLrho-g-PNIPAM core concentration was obtained, the fraction of PLLrho-g-PNIPAM recruited in the interfacial layer was calculated as being equal to the loss of the soluble polymer amount. Finally, this fraction of interfacial PLLrho-g-PNIPAM was converted in surface excess through normalization by the droplet interfacial area.

The evolution of the polymer surface excess with the droplet diameter at T = 18°C and T = 45°C is shown in **Fig. 4.C**. Experimental data were compared to the maximum reachable surface excess, corresponding to the adsorption at the interface of all the polymer chains contained in the droplets. This maximum excess increases linearly with the droplet diameter as it is proportional to the droplet volume/surface ratio. At T = 18°C, the experimental surface excess is significantly below the maximum curve and increases with droplet diameter until reaching a plateau for big droplets (> 30 µm). These observations suggest that a significant fraction of the polymer was not involved in interfacial complexation and that a saturation equilibrium between core-soluble and edge-adsorbed chains was presumably achieved at high volume/surface ratios. Interestingly, when switching temperature from T = 18°C to T = 45°C the surface excess almost doubled at a given droplet diameter. For small droplets (< 30 µm), the surface excess coincides with the maximum excess curve, indicating that all the polymer chains contained in the droplet have been segregated at the W/FCO interface. This observation is in agreement with an enrichment of the polymer interfacial layer above $T_c$, confirming the hypothesis of a core draining and an accumulation of polymer chains at the interface via hydrophobic interactions. For big droplets (> 30 µm), the experimental surface excess deviates from the maximum excess curve, suggesting that a saturation level has been reached in the SPIC process and that a fraction of soluble polymer chains could not be recruited in the interfacial complex layer.

The evolution of surface excess as a function of the polymer concentration in droplet cores is shown in **Fig. S7**. From T = 18°C to T = 45°C, the fraction of core-soluble polymer chains decreased from 0.4-0.6 to less than 0.1 concomitantly to an increase in the mean surface excess, confirming that (i) a high proportion of polymer chains were not involved in interfacial complexation in conditions of conventional SPIC (at $T < T_c$) and (ii) a massive accumulation of polymer chains at the interface was observed beyond the collapse transition (at $T > T_c$). In addition, this process seems to be mainly irreversible as the surface excess remained at the same high level when a sample was cooled back at T = 18°C after a first heating step at T = 45°C (**Fig. S8**). This suggests that the enrichment of the interfacial polymer layer is not subject to significant redispersion as PNIPAM chains are brought back to $T < T_c$.

The same quantitative approach was performed on a non-thermoresponsive PLL-g-PEG labeled with carboxy-X-rhodamine (PLLrho-g-PEG) and the results are shown in **Fig. S9**. The curves of surface excess superimpose at T = 18°C and T = 45°C and plateau at a saturation level which is twice lower than PLL-g-PNIPAM at T = 18°C (and 4 times lower at T = 45°C) indicating that (i) PLL-g-PNIPAM has a higher affinity with the interface than PLL-g-PEG presumably because PNIPAM strands are shorter than PEG ones (18 vs 44 monomers/chain respectively) and (ii) there is no effect of the temperature on the distribution of PLL-g-PEG in droplets, confirming that the process observed for PLL-g-PNIPAM is intrinsic to the thermoresponsive nature of PNIPAM strands.

*3.4. Capture of nanoparticles in the polymer interfacial layer*

The combination of nanoparticles with PLL-g-PNIPAM was investigated in order to get multicomponent interfacial layers. As a proof of concept, 50 nm NeutrAvidin-coated fluorescent nanoparticles (fluoNPs) were added in the aqueous phase containing PLL-g-PNIPAM prior to emulsification. **Fig. 5.A** show confocal micrographs of a W/FCO emulsion with 2 wt% of PLL-g-PNIPAM and 0.025 wt% of fluoNPs in the aqueous phase at T = 18°C and T = 45°C. As PLL-g-PNIPAM was not dye-labeled, the fluorescence signal was only due to fluoNPs and allowed to visualize their distribution in water droplets. At T = 18°C, the fluoNPs were homogeneously distributed within the droplets. At T = 45°C, the fluoNPs were essentially located at the droplet edge, indicating that they had been segregated at the water/FCO interface concomitantly with the collapse transition of PNIPAM at $T_c$. This process of thermo-induced co-segregation of polymer chains and fluoNPs at the interface is depicted in **Fig. 5.B** (Krytox molecules and PLL positive charges have been removed for more readability).

The evolution of the emulsion was recorded over time while temperature was increased (from 27°C to 45°C) to assess the dynamics of fluoNPs within the droplets (see Supplementary Movie). For the representative droplet showed in **Fig. 5.C**, the evolution of the radial profile and the radial kymograph are represented along the temperature sweep. In the radial kymograph, the projections of the radial fluorescence image at increasing temperatures have been put side by side in a continuous image. It illustrates the progressive disappearance of fluorescence in the droplet core concomitant with an accumulation of the fluorescence at the droplet edge. This process is also highlighted by the evolution of the radial profile as temperature rises: the uniform profile of fluorescence at low temperature ($T < T_c$) was progressively replaced by a sharp peak of fluorescence localized at the droplet edge ($T > T_c$). The consistency of the so-formed polymer-NPs mixed layer was assessed by two qualitative observations: (i) micron-size aggregates of fluoNPs that have reached the interface were totally immobilized at $T = 45°C$ as shown in **Fig. S10**, (ii) the corona of fluoNPs formed at $T = 45°C$ did not redisperse as the emulsion was cooled down below $T_c$ (**Fig. 5.D**), suggesting that the co-segregation of PLL-g-PNIPAM and fluoNPs at the interface is irreversible. These data are suggesting the formation of a highly viscous or gel-like shell at the water/FCO interface, of interest as precursors of rigid polymer capsules.

A control emulsion was made with 2wt% of non-thermoresponsive PLL-g-PEG and 0.025 wt% of fluoNPs in the aqueous phase. The distribution of fluoNPs was uniform in water droplets at $T = 18°C$ and $T = 45°C$ (**Fig. S11**), suggesting that the interfacial capture of fluoNPs observed with PLL-g-PNIPAM is due to the thermo-responsiveness of PNIPAM strands and is not inherent to the nature of fluoNPs.

*3.5. Encapsulation of a model protein*

The possibility to encapsulate proteins in PLL-g-PNIPAM core-shell droplets was tested using eGFP (enhanced Green Fluorescent Protein) as a fluorescent model protein. An aqueous solution containing 2 wt% of coumarin-labeled PLL-g-PNIPAM and 10 µM of eGFP was prepared in phosphate buffer (pH = 7.3) and emulsified in FCO phase. Polymer and protein distribution within droplets were determined by CLSM (**Fig. 6**). The bright coumarin fluorescence signal at the droplet edge shows that PLL-g-PNIPAM was concentrated at the water/FCO interface as expected, demonstrating that the presence of the protein in the aqueous phase did not affect the formation of the interfacial polymer layer induced by SPIC. The uniform distribution of eGFP within droplets showed that the protein had been efficiently encapsulated in the aqueous droplet cores without significant loss of fluorescence properties (**Fig. S12**). The overlay of confocal images clearly shows the "core-shell" distribution of the species, with the polymer at the droplet edge and the protein entrapped in the droplet core.

## 4. Conclusion

In this work, we demonstrated the possibility to enhance the formation of polymer shells by interfacial complexation in water-in-fluorocarbon oil (W/FCO) emulsions by using the thermoresponsive properties of poly(N-isopropylacrylamide) (PNIPAM) strands. A polymer layer made of poly(L-lysine)-g-poly(N-isopropylacrylamide) comb-like chains (PLL-g-PNIPAM) was formed at the edge of water droplets upon interfacial complexation between cationic groups of poly(L-lysine) and an oil-soluble anionic surfactant (Krytox). The thermoresponsive behavior of PNIPAM strands, switching from hydrophilic coils to hydrophobic globules at a transition temperature of 34°C, was used to trigger the collapse and aggregation of polymer chains upon gentle heating. Above the temperature of collapse transition, we observed a massive clustering of PLL-g-PNIPAM at the interface, likely due to hydrophobic interactions between PNIPAM strands. This process contributed to enrich the interfacial polymer layer and finally resulted in polymer:surfactant complex shells that are likely thicker than ultrathin polymer capsules fabricated by conventional interfacial complexation in W/FCO emulsion[22]. This thermo-enhancement of conventional interfacial complexation was leveraged to achieve straightforward formation of polymer-particles mixed shells. Water-soluble fluorescent nanoparticles were irreversibly segregated in the interfacial layer of PLL-g-PNIPAM after heating above the transition temperature and cooling down, suggesting that a gel-like polymer-particles shell had been formed at the droplet edge. In addition, this polymer capsular structure turned out to be an appropriate system for the encapsulation of a model protein in physiological conditions and constitutes a first step towards the design of functional capsules.

**Abbreviations:** ANS = 8-anilino-1-naphthalenesulfonic acid, CLSM = confocal laser scanning microscopy, coum = 7-(diethylamino)coumarin, eGFP = enhanced Green Fluorescent Protein, FCO = fluorocarbon oil, fluoNPs = NeutrAvidin-coated fluorescent nanoparticles, NHS = N-hydroxysuccinimidyl, PEG = poly(ethylene glycol), PLL = poly(L-lysine), PNIPAM = poly(N-isopropylacrylamide), rho = carboxy-X-rhodamine, SPIC = surfactant-polymer interfacial complexation, W/FCO emulsion = water-in-fluorocarbon oil emulsion.

**Author Contributions**

L.S., C.T., and E.M. conceived the project. L.S. prepared the samples, performed the experiments, analyzed the data, and drafted the manuscript. All authors discussed the results and contributed to the final manuscript.

**Declaration of Competing Interest**

The authors declare that they have no known competing financial interests or personal relationships that could have appeared to influence the work reported in this paper.

**Acknowledgments**

We would like to acknowledge R. Chouket, T. Le Saux and L. Jullien for providing eGFP and advice for fluorescence quantification; W. Urbach, N. Taulier and J. Fattaccioli for fruitful discussions on fluorinated species. This work was supported by the LabEx program 'DYNAMO' (ANR-11-LABEX-0011-01); the ANR grants GeneCap (ANR-17-CE09-0007) and CASCADE (ANR-17-CE09-0019); and Sorbonne University ('Ecole Doctorale 388' scholarship).

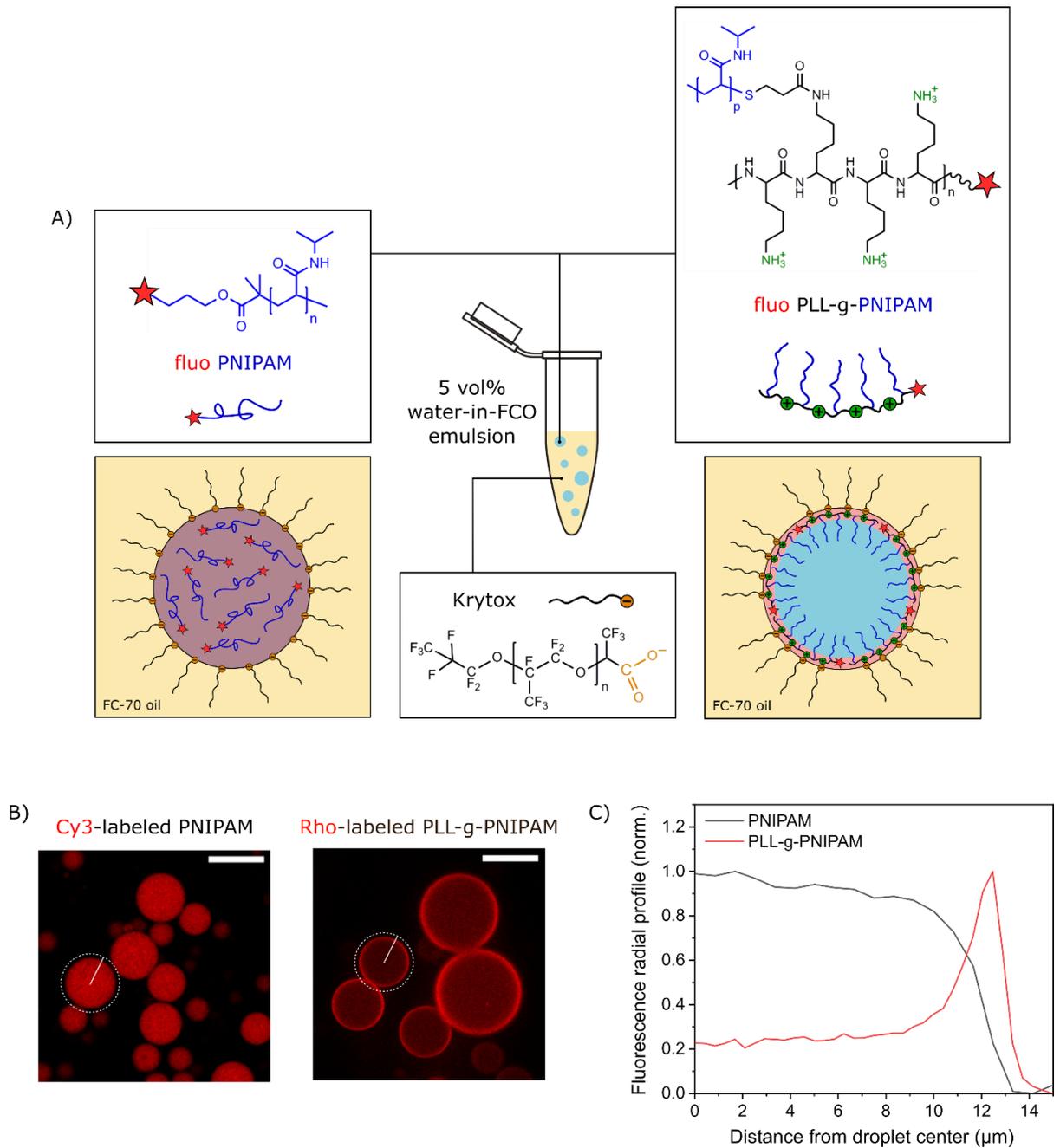

**Fig. 1.** Formulation of water-in-FCO (W/FCO) emulsions with free PNIPAM chains or PLL-g-PNIPAM copolymer in the aqueous phase and Krytox surfactant in the FCO phase. A) Schematics of the W/FCO emulsion and the expected distribution of polymer species in water droplets. B) Confocal micrographs of W/FCO emulsions with 2.5 wt% of Cy3-labeled free PNIPAM chains (left) or 3 wt% of rhodamine-labeled PLL-g-PNIPAM (right) in the aqueous phase (scale = 20 µm). C) Fluorescence radial profiles of the droplets encircled in white in the micrographs.

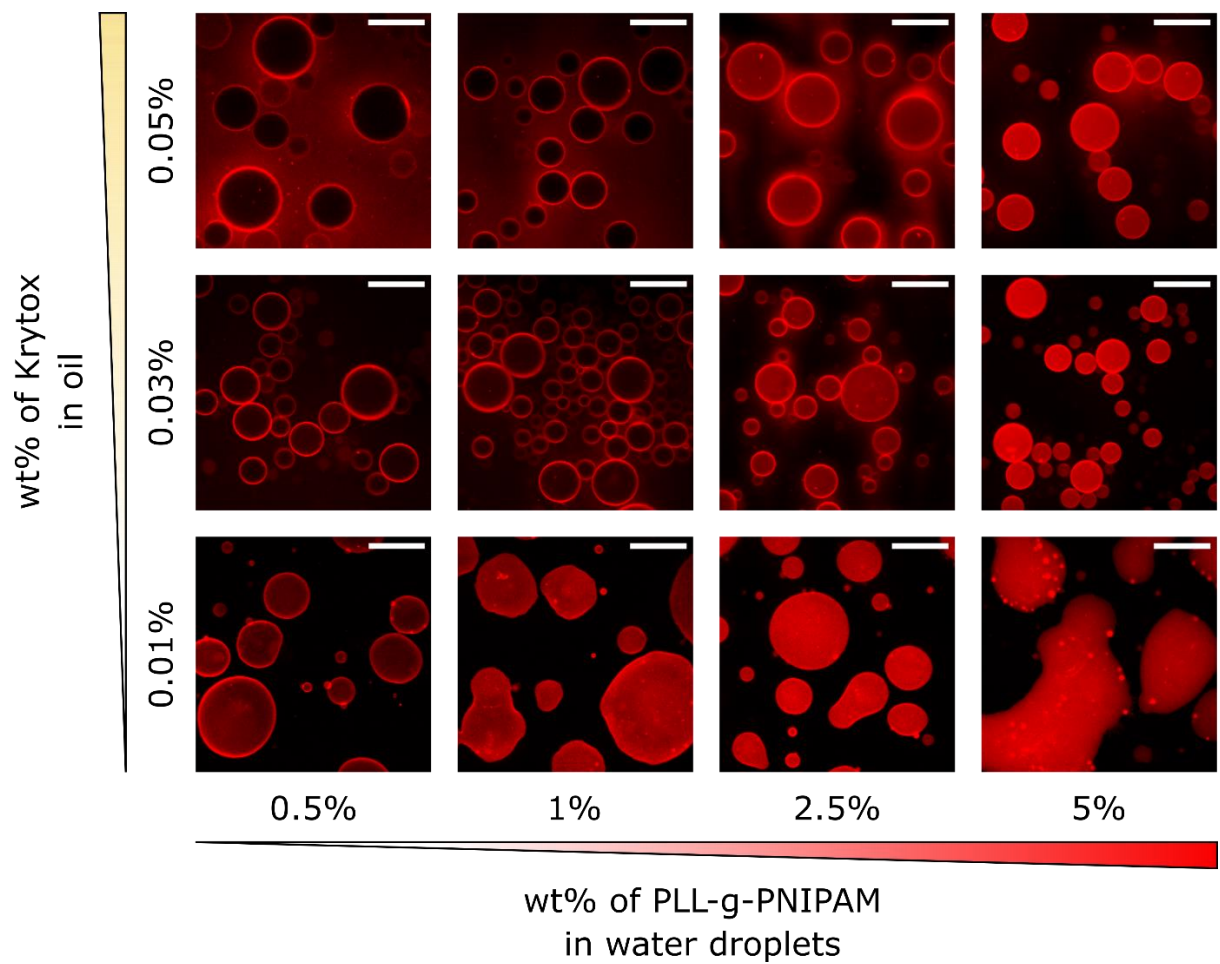

**Fig. 2**. Partial formulation diagram of water-in-FCO emulsions injected in a glass capillary and imaged by CLSM (scale bar = 50 µm). PLL-g-PNIPAM was labeled with coumarin for fluorescence imaging.

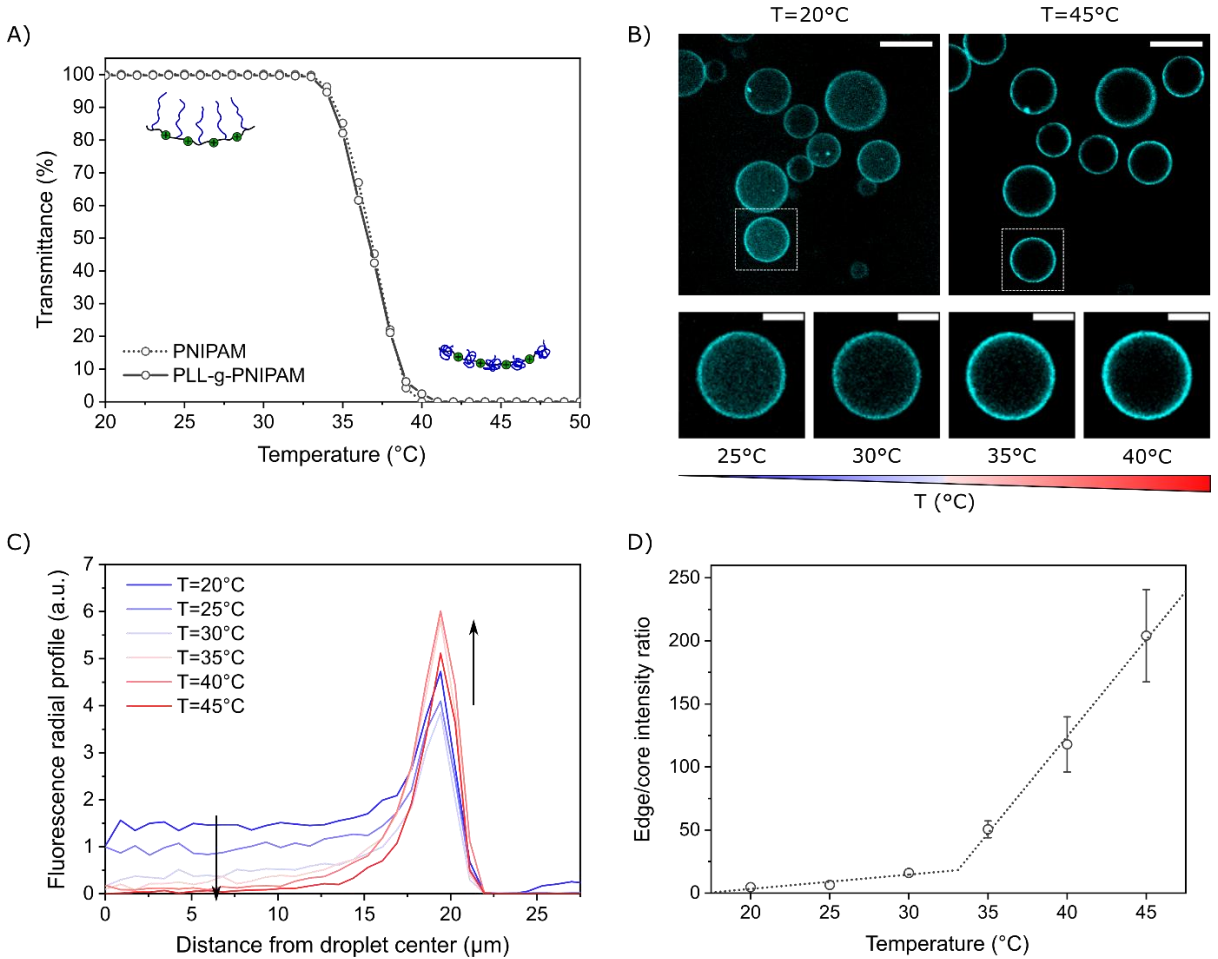

**Fig. 3.** Evidence for thermal collapse transition of PLL-g-PNIPAM in SPIC conditions. A) Turbidimetry measurement of free PNIPAM (dashed line) and PLL-g-PNIPAM (solid line) solutions at 1 wt% in phosphate buffer. B) Top: Confocal micrographs at T = 20°C and T = 45°C of a water-in-FCO emulsion with 2 wt% of PLL-g-PNIPAM and 500 µM of ANS in the aqueous phase and 0.05 wt% of Krytox in the FCO phase (scale bars = 50 µm). Bottom: Zoom on the droplet framed in white in the top micrographs, at intermediary temperatures (scale bars = 10 µm). C) Fluorescence radial profiles of the droplet shown in B at increasing temperature. D) Plot of the droplet edge/core intensity ratio - extracted from fluorescence radial profiles - as a function of the temperature (error bars = standard error on 8 droplets).

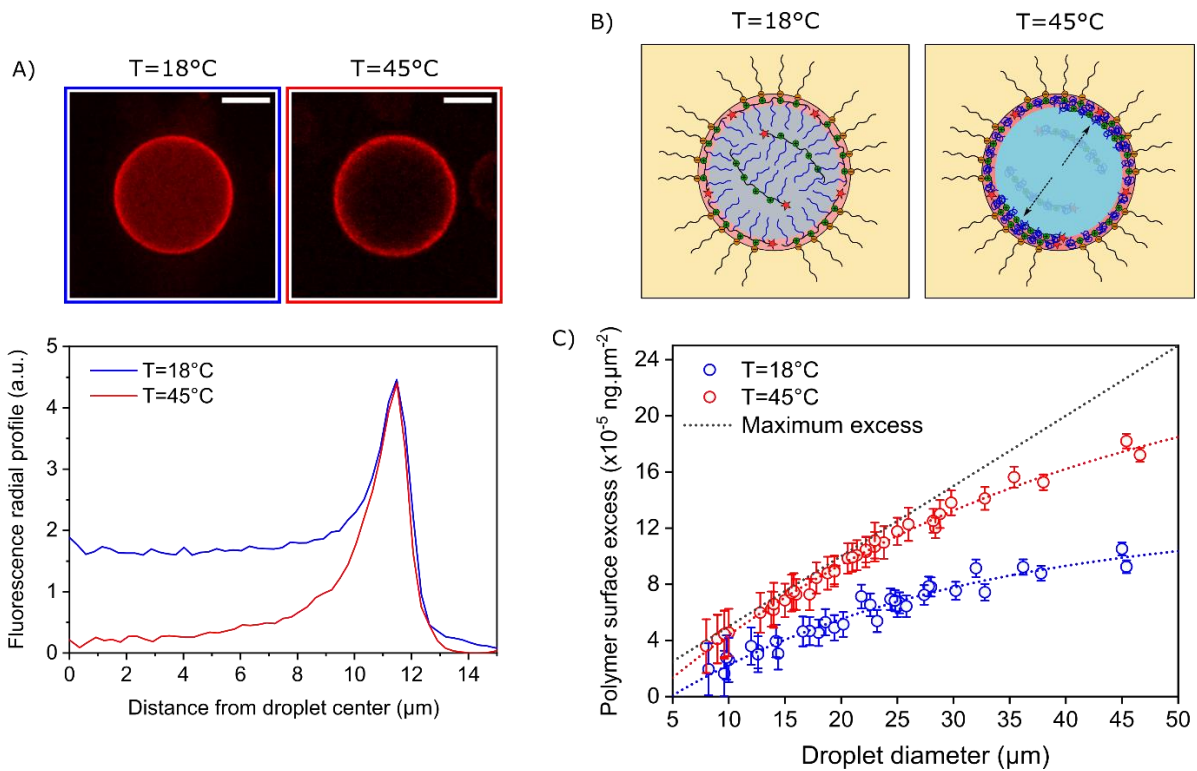

**Fig. 4.** Effect of temperature on the amount of interfacial PLLrho-g-PNIPAM. A) Top: Confocal micrographs of a representative water droplet containing 3 wt% of PLLrho-g-PNIPAM dispersed in FCO containing 0.05 wt% of Krytox at T = 18°C and T = 45°C (scale bar = 10 µm). Bottom: Fluorescence radial profiles of the top droplets. B) Schematic illustration of the accumulation of core-soluble PLLrho-g-PNIPAM chains at the interface upon collapse transition (occurring at $T_c$ = 34°C) in a water-in-FCO emulsion. C) Evolution of the surface excess of PLLrho-g-PNIPAM as a function of the droplet diameter at T = 18°C and T = 45°C, calculated from fluorescence measurements. The maximum surface excess (black dashed line) corresponds to the ideal case of interfacial adsorption of all the PLL-g-PNIPAM chains initially present in the droplet.

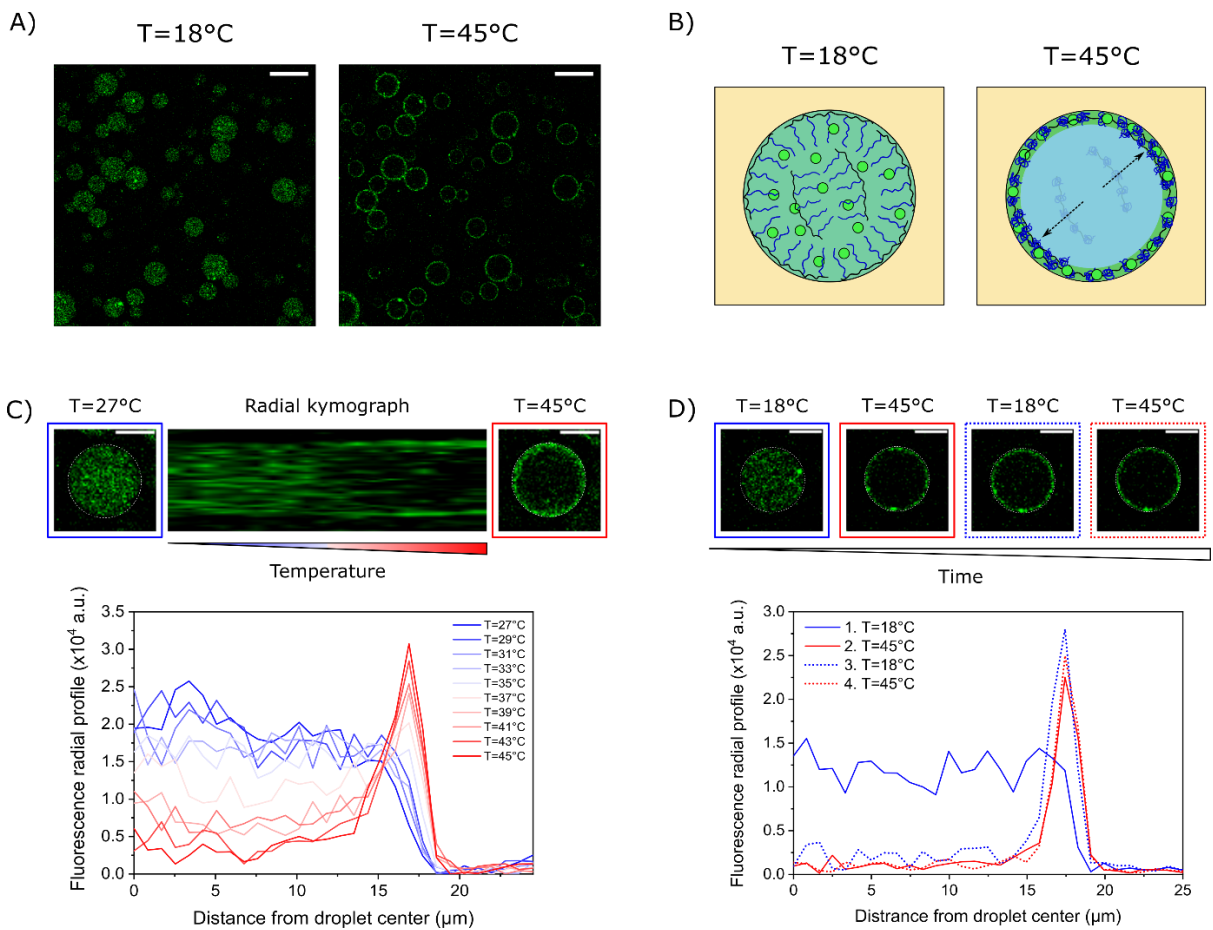

**Fig. 5.** Irreversible segregation of nanoparticles in the polymer shell above $T_c$. A) Confocal micrographs of a water-in-FCO emulsion with 2 wt% of PLL-g-PNIPAM (non-fluorescent) and 0.025 wt% of 50 nm NeutrAvidin-coated fluorescent nanoparticles (fluoNPs) in water and 0.05 wt% of Krytox in FCO, at T = 18°C and T = 45°C (scale bars = 50 µm). B) Schematic representation of the thermo-induced co-segregation of PLL-g-PNIPAM and fluoNPs at the interface upon heating. For more readability, positive charges of PLL and Krytox molecules are not represented. C) Evolution of the fluorescence signal of fluoNPs in a representative droplet over temperature rise. Top: Confocal images of the droplet at T = 27°C and T = 45°C (scale bars = 10 µm) and corresponding radial kymograph over temperature rise. Bottom: Radial fluorescence profiles of the droplet at increasing temperature. D) Confocal micrographs of a droplet over temperature switches between T = 18°C and T = 45°C and the corresponding fluorescence radial profiles.

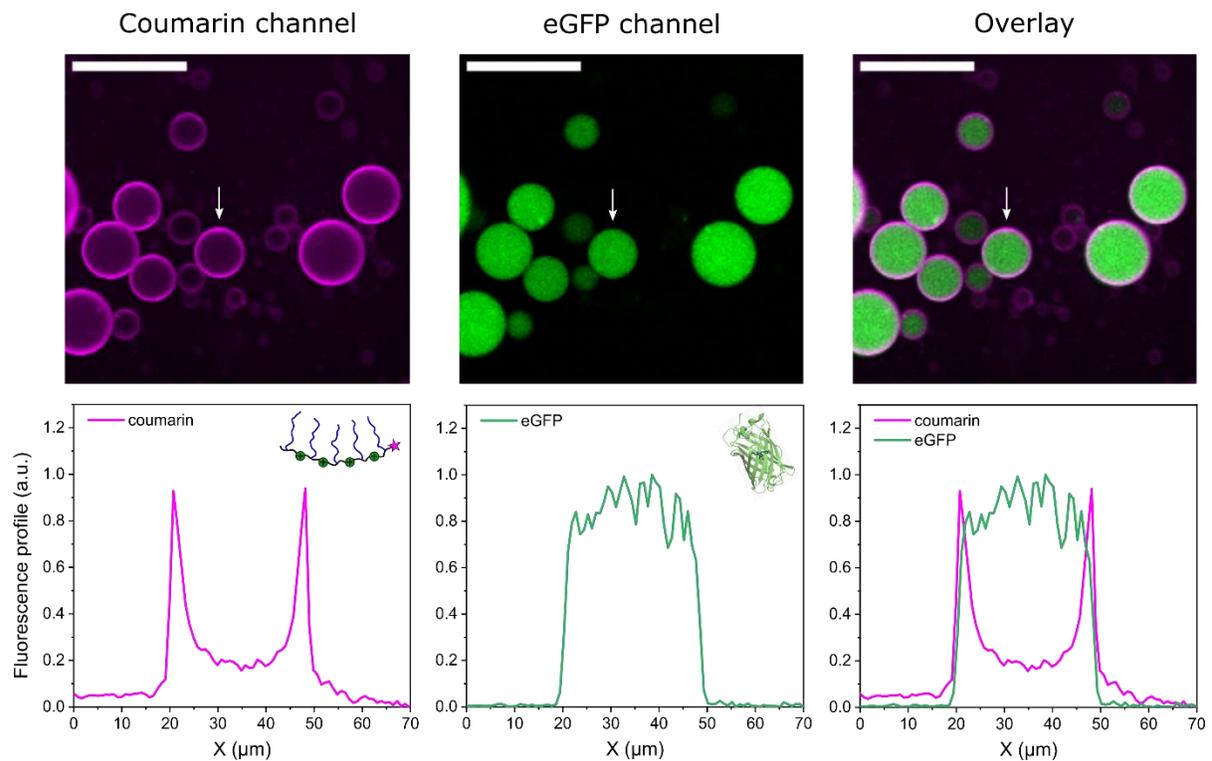

**Fig. 6.** Encapsulation of eGFP in PLL-g-PNIPAM capsule precursors. Top: Confocal micrographs of a water-in-FCO emulsion with 1 wt% of PLLcoum-g-PNIPAM and 10 µM of eGFP in the aqueous phase and 0.03 wt% of Krytox in the FCO phase at T = 18°C (scale bars = 50 µm). Bottom: Fluorescence profiles of the droplet indicated by a white arrow in the micrographs.

# Supplementary information

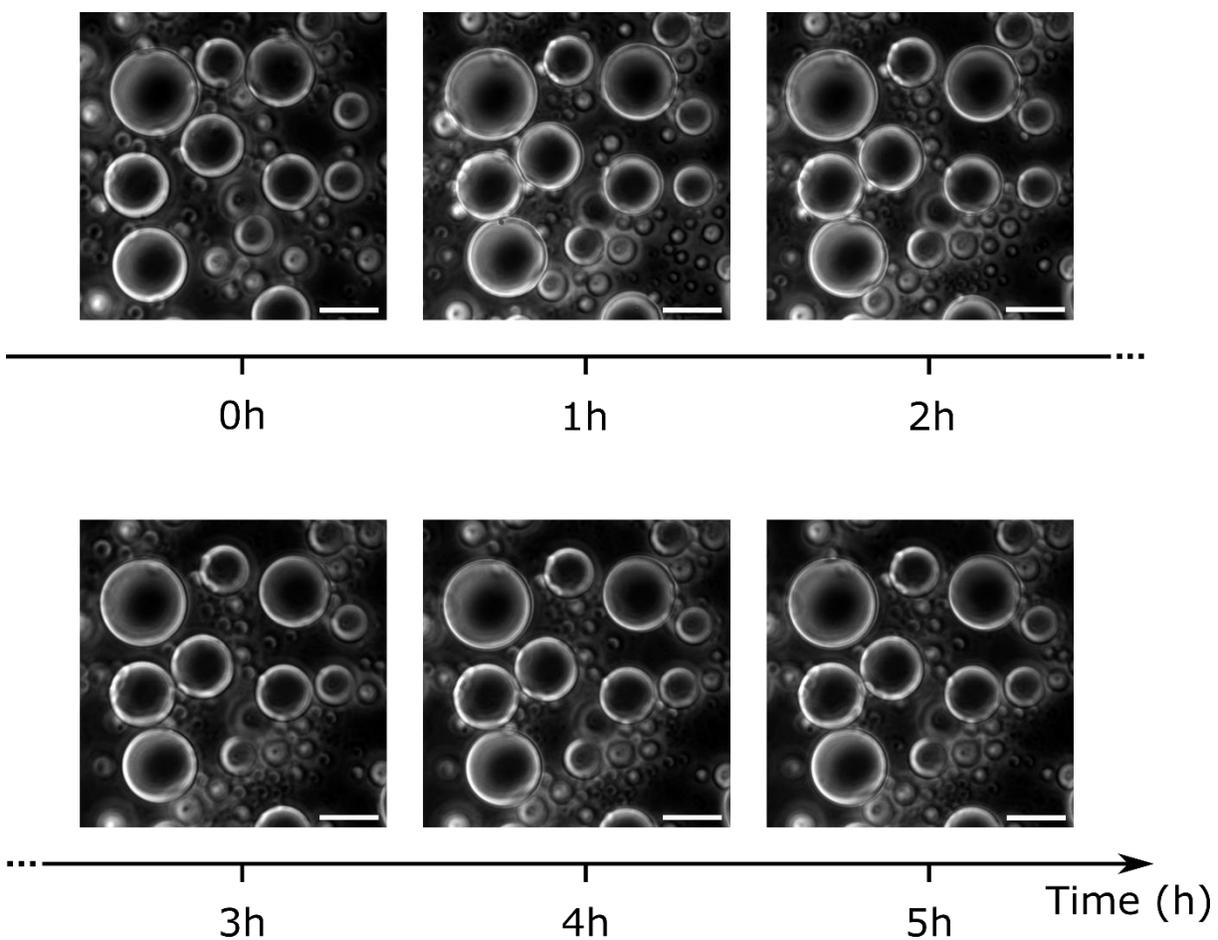

**Fig. S1**. Phase contrast microscopy images over time of a water-in-FCO emulsion with 2 wt% of PLL-g-PNIPAM in the aqueous phase and 0.05 wt% of Krytox in the FCO phase (scale bar = 30 µm). The emulsion has been injected in a glass capillary for microscopy imaging at T = 25°C.

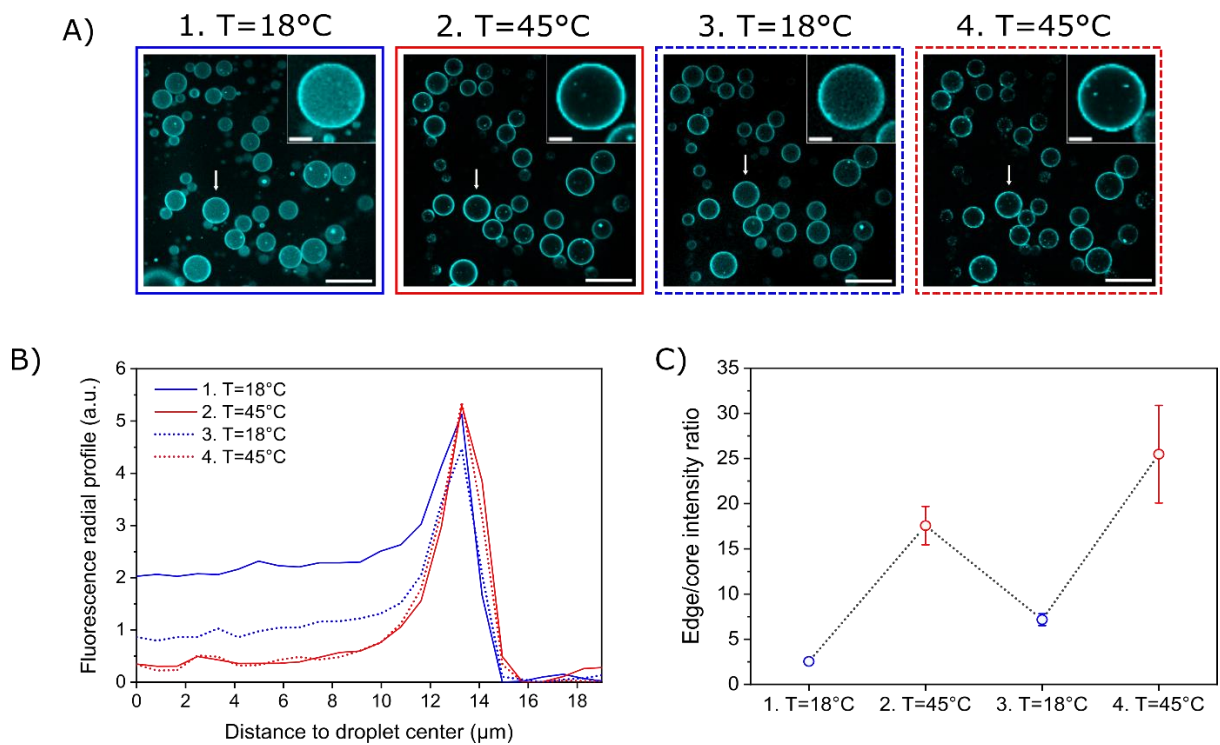

**Fig. S2**. Assessment of reversibility of ANS fluorescence in presence of PLL-g-PNIPAM. A) CLSM micrographs of a W/FCO emulsion with 2 wt% of PLL-g-PNIPAM and 500 µM of ANS in the aqueous phase and 0.05 wt% of Krytox in the FCO phase (scale bars = 50 µm). The emulsion was alternatively thermalized at T = 18°C (steps 1 and 3, blue frames) and T = 45°C (steps 2 and 4, red frames). Inserts: Zoom on the droplet indicated with a white arrow in the micrograph (scale bars = 10 µm). B) Fluorescence radial profiles of the droplet shown in inserts in A (same color code). C) Edge/core fluorescence ratio calculated from the curves in B (same color code, error bars = standard error on 8 droplets).

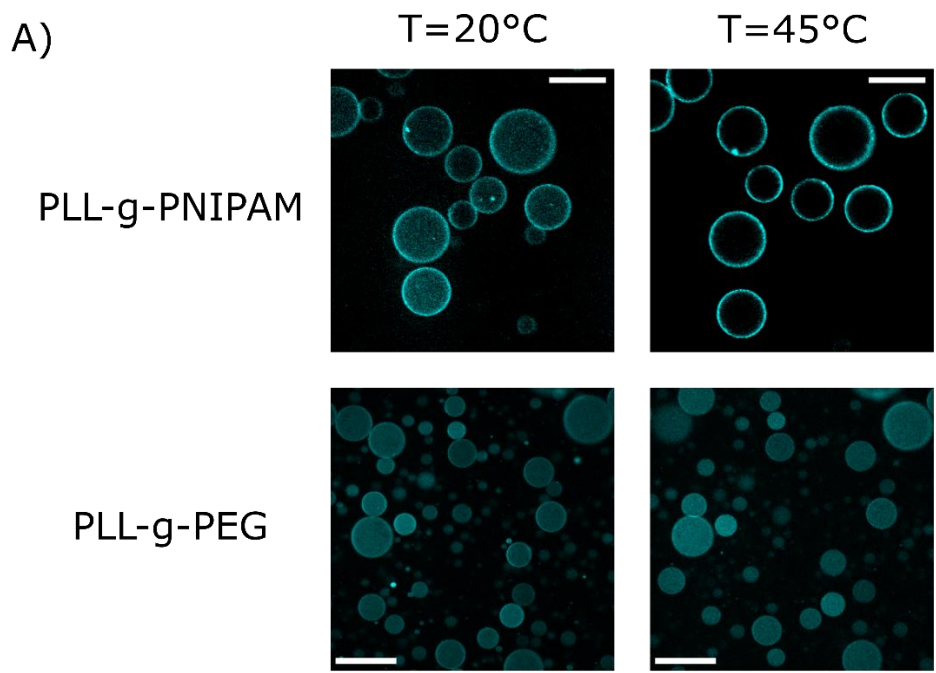

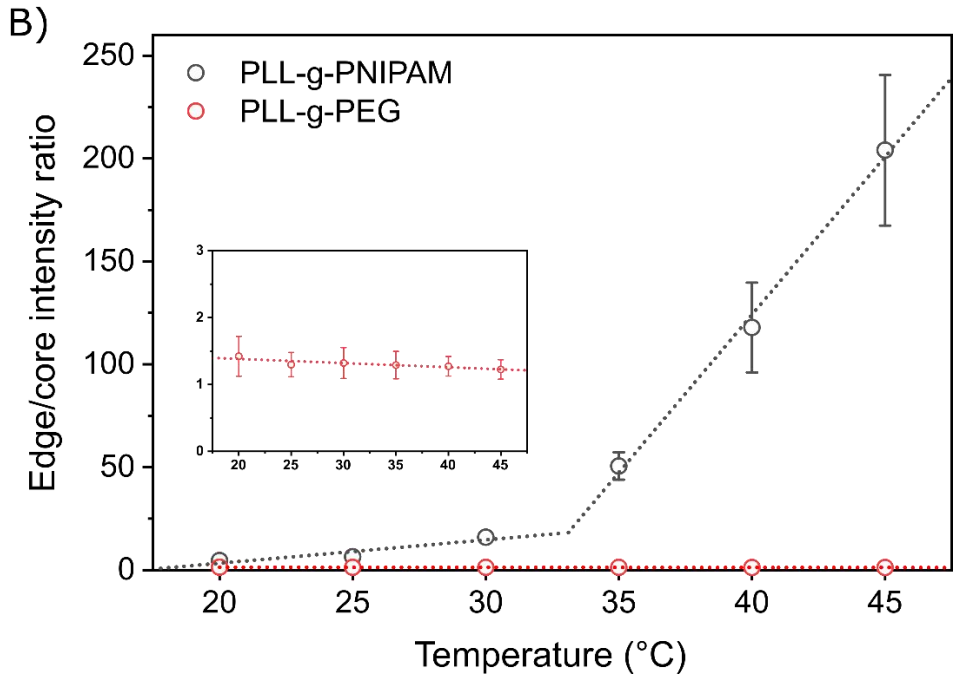

**Fig. S3.** Comparison of PLL-g-PNIPAM and PLL-g-PEG regarding ANS fluorescence. A) CLSM micrographs of a W/FCO emulsion with 2 wt% of PLL-g-PNIPAM or PLL-g-PEG and 500 µM of ANS in the aqueous phase, and 0.05 wt% of Krytox in the FCO phase, at T = 20°C and T = 45°C (scale bars = 50 µm for PLL-g-PNIPAM and 30 µm for PLL-g-PEG). B) Edge/core intensity ratio calculated from the radial fluorescence profile of 8 droplets (error bars = standard error) plotted as a function of the temperature for PLL-g-PNIPAM (black) and PLL-g-PEG (red).

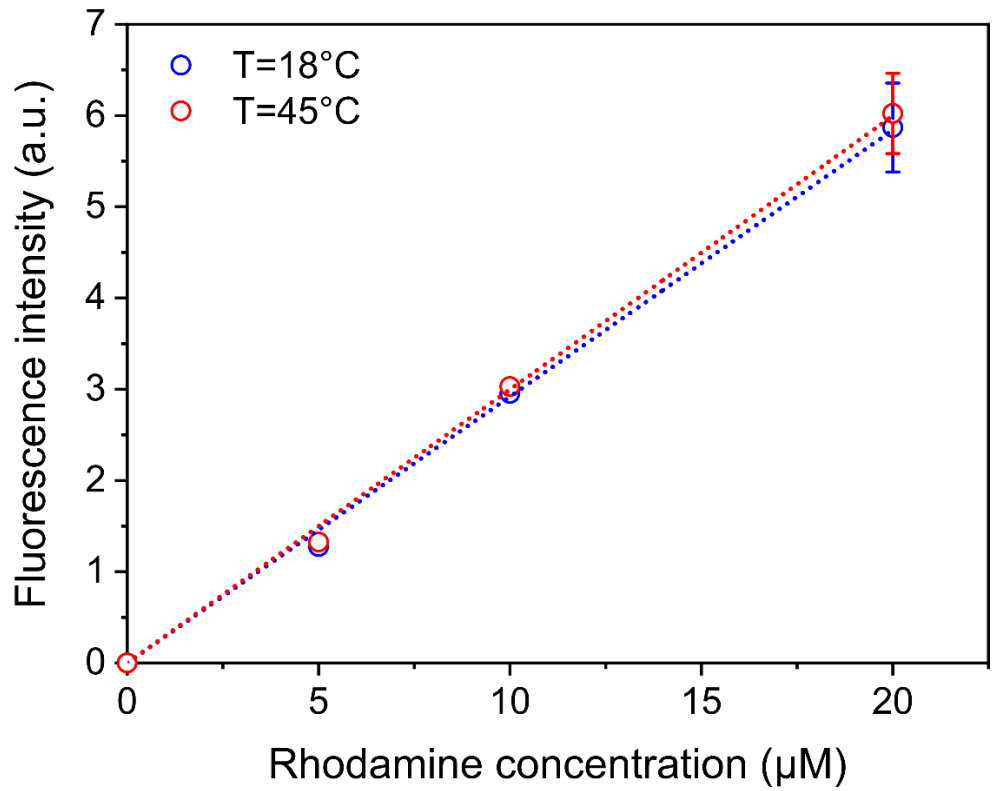

**Fig. S4.** Fluorescence intensity of carboxy-X-rhodamine (rho) solutions measured with a confocal microscope in the same irradiation conditions at T = 18°C and T = 45°C.

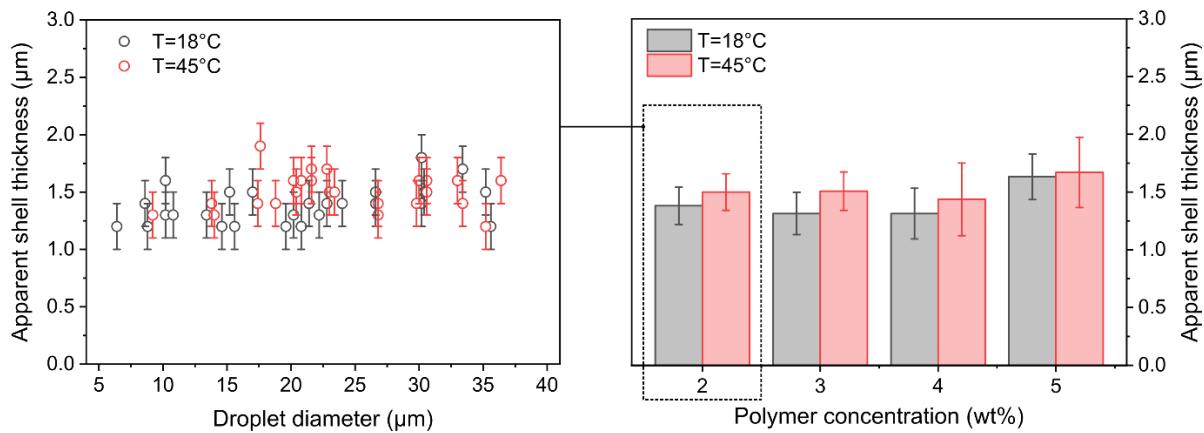

**Fig. S5.** Left: Fluorescent shell thickness measured on a W/FCO emulsion with 2 wt% of PLLrho-g-PNIPAM in the aqueous phase and 0.05 wt% of Krytox in the FCO phase at T = 18°C and T = 45°C. Right: Mean shell thickness measured on emulsions with PLLrho-PNIPAM concentration in the range 2-5 wt% (error bars = standard deviation).

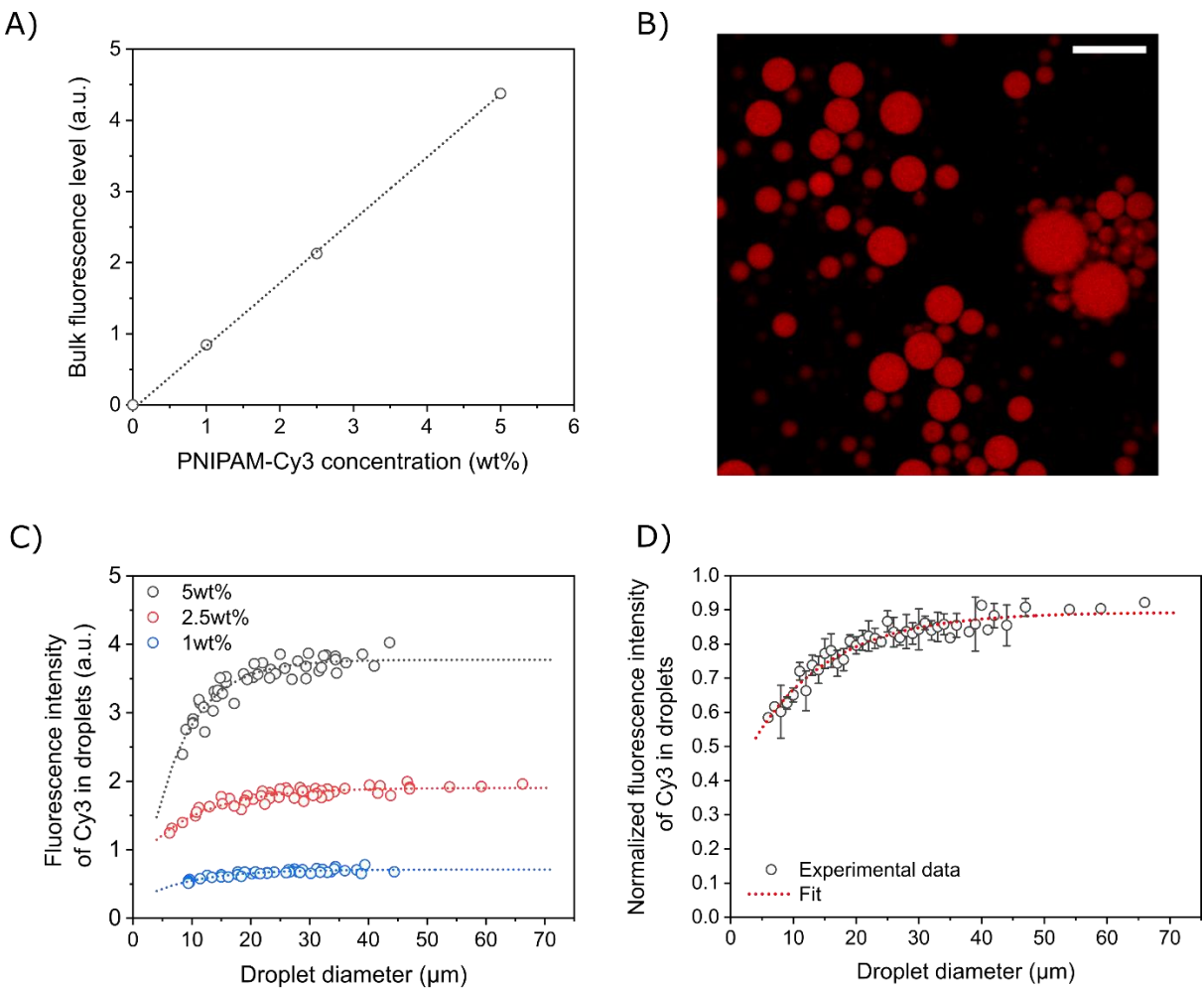

**Fig. S6.** A) Calibration curve of PNIPAM-Cy3 fluorescence measured in bulk solution by CLSM. B) Confocal micrograph of a W/FCO emulsion with 2.5 wt% of PNIPAM-Cy3 in water and 0.05 wt% of Krytox in FCO at T = 18°C (scale bar = 50 µm). C) Fluorescence level of PNIPAM-Cy3 measured in the core of water droplets as the function of the droplet diameter and for different concentrations of PNIPAM-Cy3 D) Normalized fluorescence level in droplets (corresponding to the ratio between core fluorescence and bulk fluorescence at the same concentration) as a function of the droplet diameter. To simplify the distribution, each droplet diameter has been rounded down to the nearest integer (error bars = standard deviation for each integer diameter).

To calibrate the fluorescence of rhodamine-labeled PLL-g-PNIPAM (PLLrho-g-PNIPAM), optical bias related to droplet size and curvature had to be corrected. As Cy3-labeled PNIPAM (PNIPAM-Cy3) was homogeneously distributed within droplet cores, it was used as a reference. Based on the fluorescence of PNIPAM-Cy3 in droplets, we determined a corrective factor accounting for droplet size effects on the fluorescence signal. The fluorescence intensity of aqueous solutions of PNIPAM-Cy3 at various concentrations ($I_{bulk}^{Cy3}(C)$) was measured by CLSM (**Fig. S6.A**). In the same conditions of irradiation, the fluorescence intensity of PNIPAM-Cy3 ($I_{drop}^{Cy3}(C,d)$) was measured in a polydisperse W/FCO emulsion with water droplets containing PNIPAM-Cy3 at the same concentration than the calibration curve (**Fig. S6.C**). The normalization of droplet intensities by the bulk intensity at the same PNIPAM-Cy3 concentration resulted in a corrective factor $I_{norm}^{Cy3}(d) = \frac{I_{drop}^{Cy3}(C,d)}{I_{bulk}^{Cy3}(C)}$ that only depends on the droplet diameter $d$ (**Fig. S6.D**).

The same normalization approach was followed for the fluorescence of PLLrho-g-PNIPAM: the fluorescence intensity of different PLLrho-g-PNIPAM solutions was measured in bulk ($I_{bulk}^{rho}(C)$). The same solutions were emulsified in FCO containing Krytox. The fluorescence intensity of PLLrho-g-PNIPAM ($I_{drop}^{rho}(C,d)$) was then measured in the droplet cores and normalized by $I_{bulk}^{rho}(C)$.

The ratio between the actual PLLrho-g-PNIPAM concentration in droplet cores $C_{core}$ and the analytical concentration $C$ introduced in the aqueous phase before emulsification was eventually determined by the following formula:

$$\frac{C_{core}}{C} = \frac{I_{drop}^{rho}(C,d)}{I_{bulk}^{rho}(C)} \times \frac{I_{bulk}^{Cy3}(C)}{I_{drop}^{Cy3}(C,d)} = \frac{I_{norm}^{rho}(d)}{I_{norm}^{fluo}(d)}$$

The factor $I_{norm}^{rho}(d)$ evaluates the inhomogeneity of the PLLrho-g-PNIPAM distribution in water droplets due to the massive adsorption of the polymer at the water/FCO interface and the factor $I_{norm}^{fluo}(d)$ accounts for the correction of optical bias.

PLLrho-g-PNIPAM surface excess $\Gamma$ was defined as the excess number of polymer chains adsorbed at the interface in conditions of SPIC (compared to a uniform distribution) per surface unit. It is given by the following formula:

$$\Gamma = \frac{N_{excess}}{S} = \frac{(C - C_{core})V}{S} = \left(1 - \frac{C_{core}}{C}\right)\frac{d}{6C} = \left(1 - \frac{I_{norm}^{rho}(d)}{I_{norm}^{Cy3}(d)}\right)\frac{d}{6C}$$

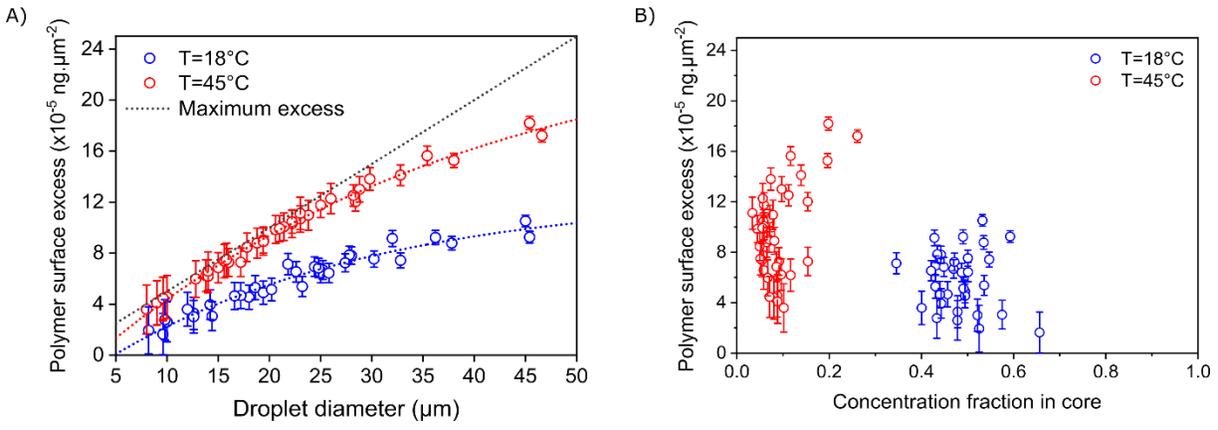

**Fig. S7.** Evolution of PLLrho-g-PNIPAM surface excess with droplet diameter (A) and concentration fraction in droplet core (B) at T = 18°C and T = 45°C. The W/FCO emulsion was made of 2 wt% of PLLrho-g-PNIPAM in the water phase and 0.05 wt% of Krytox in the FCO phase. The concentration fraction in core corresponds to the ratio between the concentration measured in the droplet cores and the analytical concentration introduced in the formulation (3 wt%).

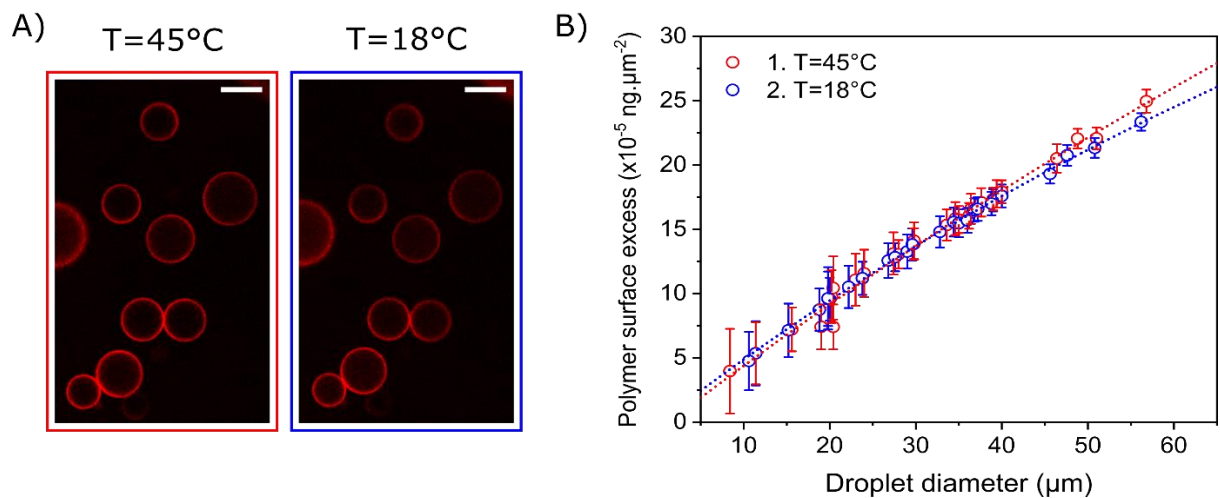

**Fig. S8**. Assessment of the reversibility of PLL-g-PNIPAM adsorption at the water/FCO interface upon collapse transition. A) Confocal micrographs of a W/FCO emulsion with 3 wt% of PLLrho-g-PNIPAM in water and 0.05 wt% of Krytox in FCO, heated up at T = 45°C and cooled back to T = 18°C (scale bars = 20 µm). Thermalization was ensured for 5min before measurement. B) PLLrho-g-PNIPAM surface excess calculated from the measurement of rhodamine fluorescence in the droplet cores at T = 45°C and T = 18°C (error bars = propagation of uncertainty).

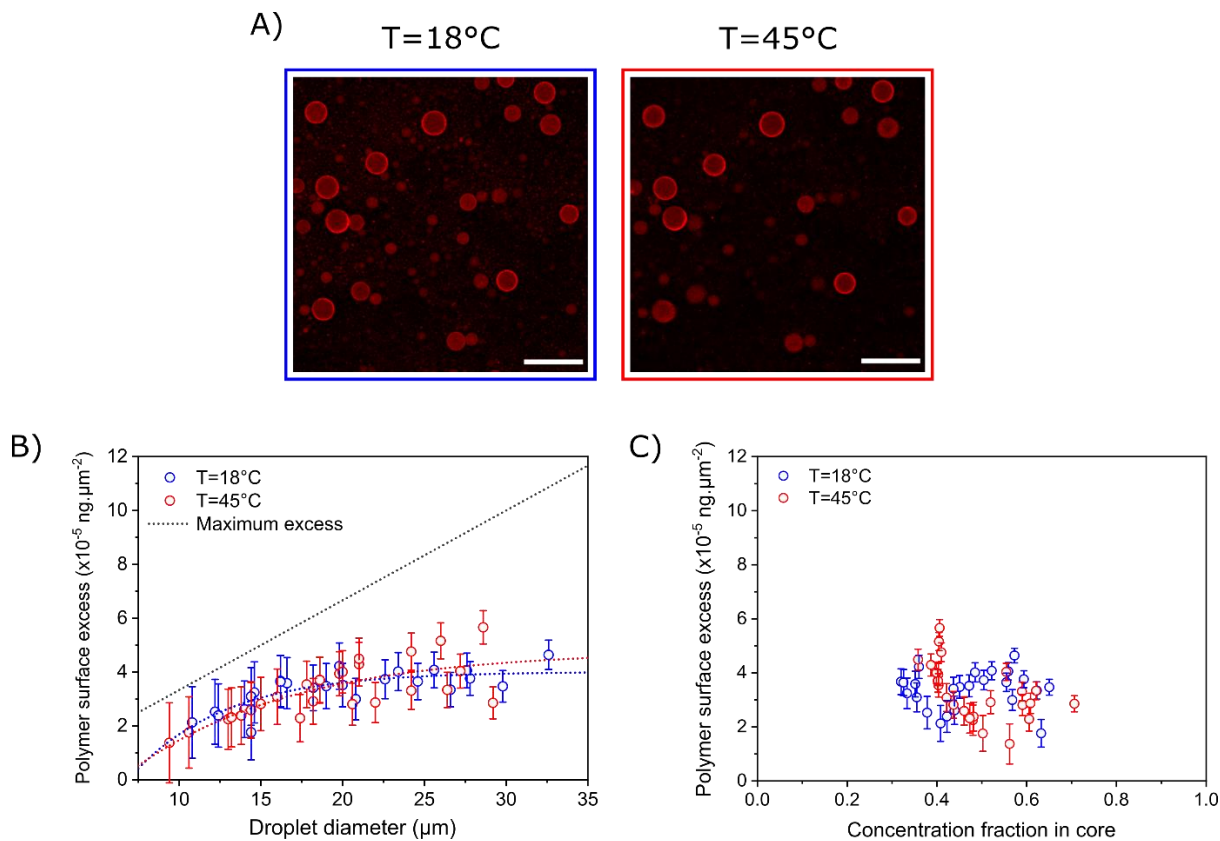

**Fig. S9.** Effect of temperature on a non-thermoresponsive PLLrho-g-PEG layer. A) Confocal micrographs of a W/FCO emulsion with 2 wt% of PLLrho-g-PEG in water and 0.05 wt% of Krytox in FCO, at T = 18°C and T = 45°C (scale bars = 20 µm). B) PLLrho-g-PEG surface excess calculated from the measurement of rho fluorescence in the droplet cores as a function of the droplet diameter at T = 18°C and T = 45°C (error bars = propagation of uncertainty). C) PLLrho-g-PEG surface excess as a function of the concentration fraction in the droplet cores at T = 18°C and T = 45°C.

**Supplementary Movie**

Evolution over temperature rise (27-45°C) of a water-in-FCO emulsion with 2 wt% of PLL-g-PNIPAM and 0.025 wt% of 50nm NeutrAvidin-coated fluorescent nanoparticles (fluoNPs) in water and 0.05 wt% of Krytox in FCO phase (scale bar = 50 µm). The displayed temperature was measured on the microscope stage supporting the sample. Confocal micrographs were acquired every 10 s and are represented at 5 fps in the movie.

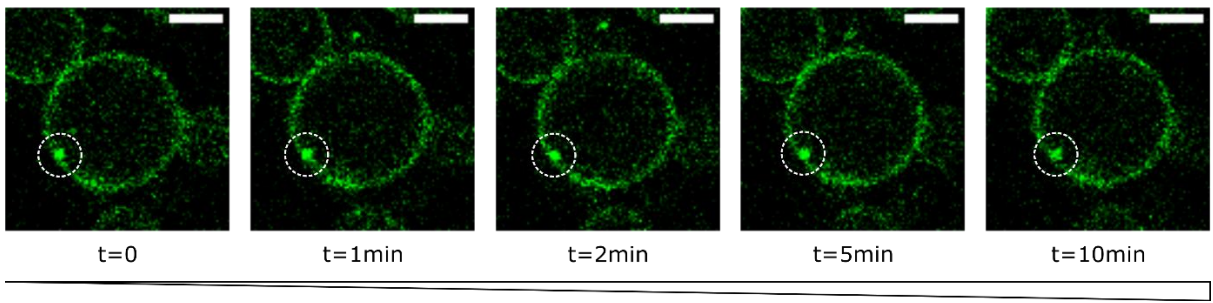

**Fig. S10**. Evolution over time at T =4 5°C of a water droplet containing 2 wt% of PLL-g-PNIPAM and 0.025 wt% of 50 nm NeutrAvidin-coated fluorescent nanoparticles (continuous phase = FCO with 0.05 wt% of Krytox, scale bars = 10 µm)

The segregation of fluoNPs at the water/oil interface has been triggered by the accumulation of PLL-g-PNIPAM towards the droplet edge upon collapse transition. The immobilization of a micrometric cluster of fluoNPs at the interface over 10 min (encircled in white in Fig. S10) tends to indicate that the mixed interfacial layer made of PLL-g-PNIPAM and fluoNPs behaves as a viscous or gel-like shell.

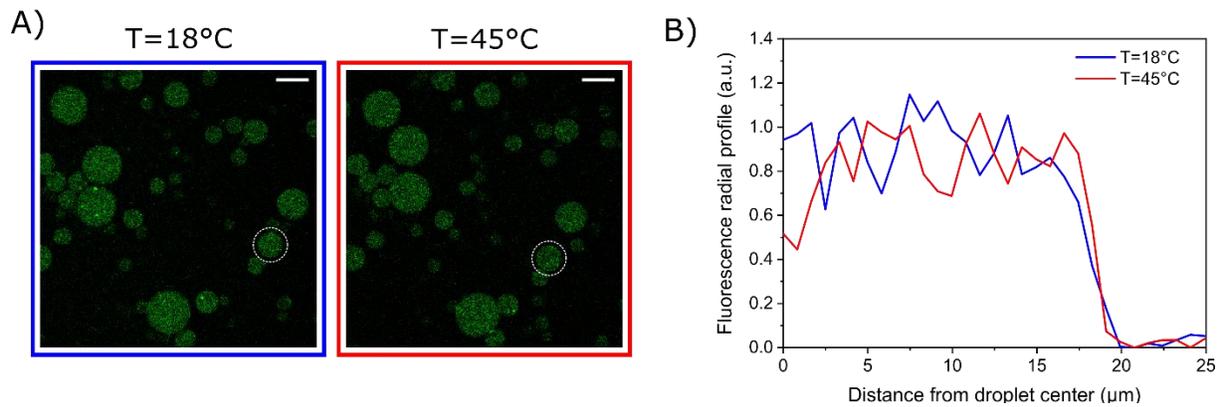

**Fig. S11.** A) Confocal micrographs of a W/FCO emulsion with 2 wt% of PLL-g-PEG (non-fluorescent) and 0.025 wt% of NeutrAvidin-coated fluorescent nanoparticles in water and 0.05 wt% of Krytox in FCO, at T = 18°C and T = 45°C (scale bars = 50 µm). B) Radial fluorescence profile of the droplet encircled in white at T = 18°C and T = 45°C.

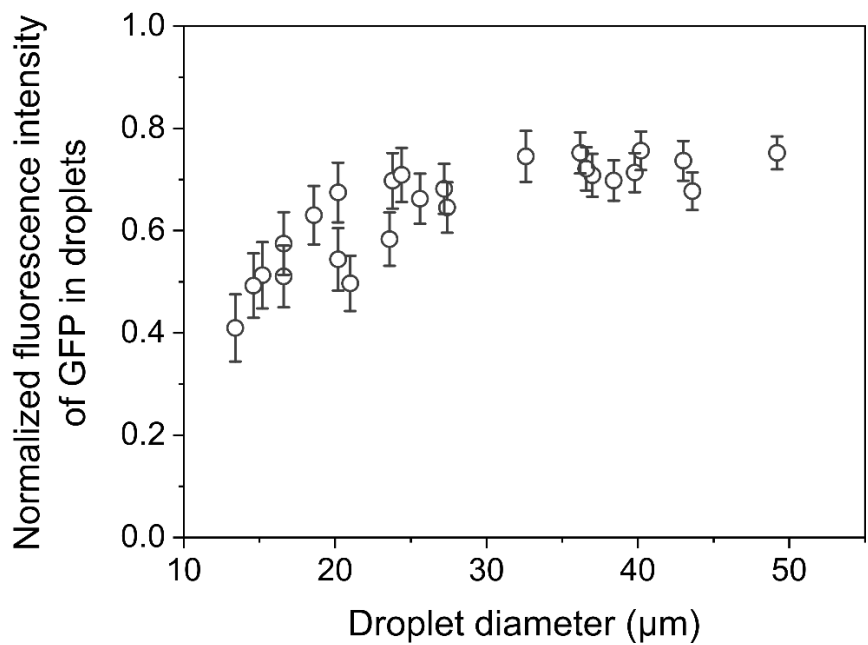

**Fig. S12**. Normalized fluorescence level of eGFP in W/FCO emulsion with 1 wt% of coumarin-labeled PLL-g-PNIPAM and 10 µM of eGFP in the aqueous phase and 0.03 wt% of Krytox in FCO.

The fluorescence level of eGFP in droplet cores has been normalized by the fluorescence level of the bulk aqueous solution and corrected by the size factor measured on PNIPAM-Cy3 emulsion as described in Fig. S7. As first approximation, the normalized fluorescence intensity of eGFP quantifies the preservation of GFP integrity upon the emulsification process. ~ 70-75% of the eGFP initial fluorescence was detected in droplets bigger than 25 µm, indicating that the protein had not been dramatically denatured during the encapsulation in water droplets. In droplets smaller than 25 µm, the higher loss of fluorescence intensity in water droplets may be reasonably attributed to optical distortion of the fluorescence signal (not compensated by the correction process) and to bleaching of eGFP, rather than to a significant alteration of eGFP properties.